\documentclass[final,1p,times]{elsarticle}

\usepackage{pifont}
\usepackage{natbib}
\usepackage{geometry}
\usepackage{graphicx}
\usepackage{txfonts}
\usepackage{subfigure}
\usepackage{wrapfig}
\usepackage{color}
\usepackage{booktabs}
\usepackage{caption}
\usepackage{amsmath}
\usepackage{amssymb}
\usepackage{url}
\usepackage{colortbl}
\usepackage{multirow}
\usepackage{fnbreak}
\usepackage{helvet}
\usepackage[font=bf]{caption}

\renewcommand{\familydefault}{\sfdefault}

\graphicspath{{figs/}}

\definecolor{Gray}{gray}{0.7}

\begin{document}

\begin{frontmatter}

\begin{center}
\small \textbf{5\textsuperscript{th} IAA Planetary Defense Conference -- PDC 2017} \\ \textbf{15--19 May 2017, Tokyo, Japan} \\ \vspace{0.1in} \textbf{IAA-PDC17-05-18} \\
\end{center}

\title{Parameter-space study of kinetic-impactor mission design}  


\author[fml1]{Alexandre Payez\corref{cor1}\fnref{fn1}}
\ead{alexandre.payez@esa.int}
\author[fml2]{Johannes Schoenmaekers\fnref{fn2}}
\ead{johannes.schoenmaekers@esa.int}

\cortext[cor1]{Corresponding author}
\fntext[fn1]{Postdoctoral research fellow, Mission Analysis Section}
\fntext[fn2]{Head of the Flight Dynamics Division}

\address[fml1]{ESA--ESOC, Robert-Bosch-Str. 5, Darmstadt, D-64293, Germany, +49 6151 90 2305}
\address[fml2]{ESA--ESOC, Robert-Bosch-Str. 5, Darmstadt, D-64293, Germany, +49 6151 90 2282}


\begin{abstract}
While almost all potentially hazardous asteroids (PHAs) with a size larger than one kilometre have been discovered, it is well-known that the vast majority of the smaller ones are in fact yet to be found.
There is therefore an excellent motivation to consider at once all possible Earth-crossing orbits, and to undertake a systematic study of mitigation missions for the entire parameter space of orbital elements.

It is shown that the whole parameter space can be reduced, without loss of generality, to only three relevant dimensionless parameters: the eccentricity and inclination of the asteroid orbit, and the asteroid true anomaly at impact.
Ballistic kinetic-impactor mitigation missions are studied for the entire parameter space, considering critical feasibility constraints such as the launcher performance and the illumination conditions at deflection.
Different classes of optimal solutions are found to exist and can be directly linked to asteroid orbital properties.
The aim of this work is to help identify an appropriate response to the potential threat of a collision of a near-Earth object with our planet, to provide a preliminary mission design, and to determine in which parts of parameter space difficulties may arise.

The problem is studied in three levels.
The first one is an analytical optimisation study which ignores the phasing ({\it i.e.} the Earth and the asteroid are always assumed to be where they are needed), and with both launch and deflection always located on the line of nodes of the asteroid.
The new parametrisation indeed leads to simpler equations, and also enables the study of daytime and nighttime impacts at once (which demand significant changes in the required strategy) since it preserves a symmetry of the impact geometry.
These results are then validated with a study of the optimised trajectories which again ignores the phasing, and finally with a full optimisation including the phasing.
In the end, the analytical results indeed prove to be useful to determine which missions can be performed, and to provide preliminary mission parameters.
Such results can moreover be used to highlight and provide more insight into the driving physical, geometrical, and launcher dependencies, while helping with the identification of problematic regions as a function of the launcher performance.
\end{abstract}



\end{frontmatter}

%
%

\section{Objective and assumptions}
\label{sec:intro}

This paper presents some of the new results that have been derived in Refs.~\cite{AP:param} and~\cite{APJS:defl}.
The objective of these works is to provide a preliminary mitigation-mission design for any asteroid that might be on a collision course with our planet\hspace{1pt}---\hspace{1pt}with the intent of mapping threat scenarios to mission types. The adopted strategy for that purpose is to consider at once all the conceivable elliptical asteroid orbits able to strictly impact Earth.
We build upon an earlier study~\cite{FB:2015} and adopt a complementary approach. These are not the first works of this type, and we therefore refer the reader in particular to Refs.~\cite{SanchezColombo:2013,ThiryVasile:2016} and the references therein for other methods and results.
Henceforth, ``PHA'' refers to asteroids with vanishing minimal orbit intersection distance (MOID $= 0$); an excellent approximation for any actual asteroid about to hit our planet.\footnote{It is necessarily tiny compared to the other scale of the problem: the typical heliocentric distance (equiv. the semi-major axis).}
For clarity, in what follows, ``impact'' ($\cdot_I$) always means the PHA--Earth encounter, while for the mitigation of the asteroid using the spacecraft, the word ``deflection'' ($\cdot_D$) is used.

Another approximation that we make is to assume the Earth motion around the Sun to be circular, thereby increasing the symmetry of the problem. For our needs, the main benefit is that neglecting the tiny eccentricity of the actual Earth orbit (of the order of $0.0167$) gives one the freedom to redefine or rotate, for any asteroid, where the impact will take place, since there is then no preferred direction on the Earth orbit  ({\it i.e.\@} the physics no longer depends on the longitude of ascending node of the asteroid orbit $\Omega$).
As in Ref.~\cite{FB:2015}, considering a circular Earth orbit of radius equal to one astronomical unit (AU), one can always choose the impact location $\vec{r}_I$ to be given for instance by
\begin{equation}
	\vec{r}_I = 1~{\rm AU} \ \vec{e}_x,
	\label{eq:rI}
\end{equation}
for any PHA that could be imagined; thus placing the PHA line of nodes along the $x$-direction.
A future strict impact with a general PHA orbit, inclined with respect to the Earth orbit, can of course only happen where the asteroid crosses the ecliptic, namely at one of its two nodes.
A judicious reference frame for the problem at hand is then for instance to have $(\vec{e}_x,\vec{e}_y)$ in the ecliptic plane with $\vec{e}_y$ parallel to the velocity of the Earth at $\vec{r}_I$, and the positive $z$-direction along its angular-momentum vector.

\section{A new natural parametrisation for the impact problem}

Since the aim is to study the whole parameter space of ballistic kinetic-impactor missions to Earth-crossing asteroids, finding a suitable parametrisation to describe any such asteroid is our first objective.

\subsection{Starting point: restoring a symmetry of the impact geometry}

\label{sec:branches}

First of all, it is worth realising that the relative geometry of the impact problem is independent of whether the impact location corresponds to the ascending or to the descending node of the PHA orbit: it remains completely identical whether the asteroid is going to hit the Earth from above or from below.
When considering fictitious asteroids, we therefore want to avoid having to specify explicitly at which node the impact will take place, since doing otherwise would actually hide a symmetry of the problem.

Now, as well-known, for any fixed semi-major axis $a$, eccentricity $e$, and inclination $i$ such that an impact could be possible, there are two truly distinct PHA orbits, oriented differently: either such that the PHA crosses the Earth orbit from the inside to the outside, or {\it vice versa}, as illustrated in Fig.~\ref{fig:daytimenighttime}. Since at close approach the asteroid correspondingly moves either towards the illuminated side or towards the dark side of the Earth, one can for instance respectively talk about \emph{daytime} and \emph{nighttime} impacts~\cite{FB:2015}.

The distinction between those two branches can for instance be done via their respective argument of perihelion $\omega$ (counted from the ascending node, by definition). 
However, doing so of course breaks the symmetry that was just discussed. 
Not only does this requires explicitly choosing the impact node, using $\omega$ furthermore puts an extra emphasis on the ascending node, which is ill-suited and arbitrary.

It is however possible to restore that symmetry, and doing so bears great benefits.

\subsection{Solution: use the asteroid true anomaly at impact}
\label{sec:introducing_fI}

\begin{figure}[h!]
    \begin{center}
        \includegraphics[height=7.cm]{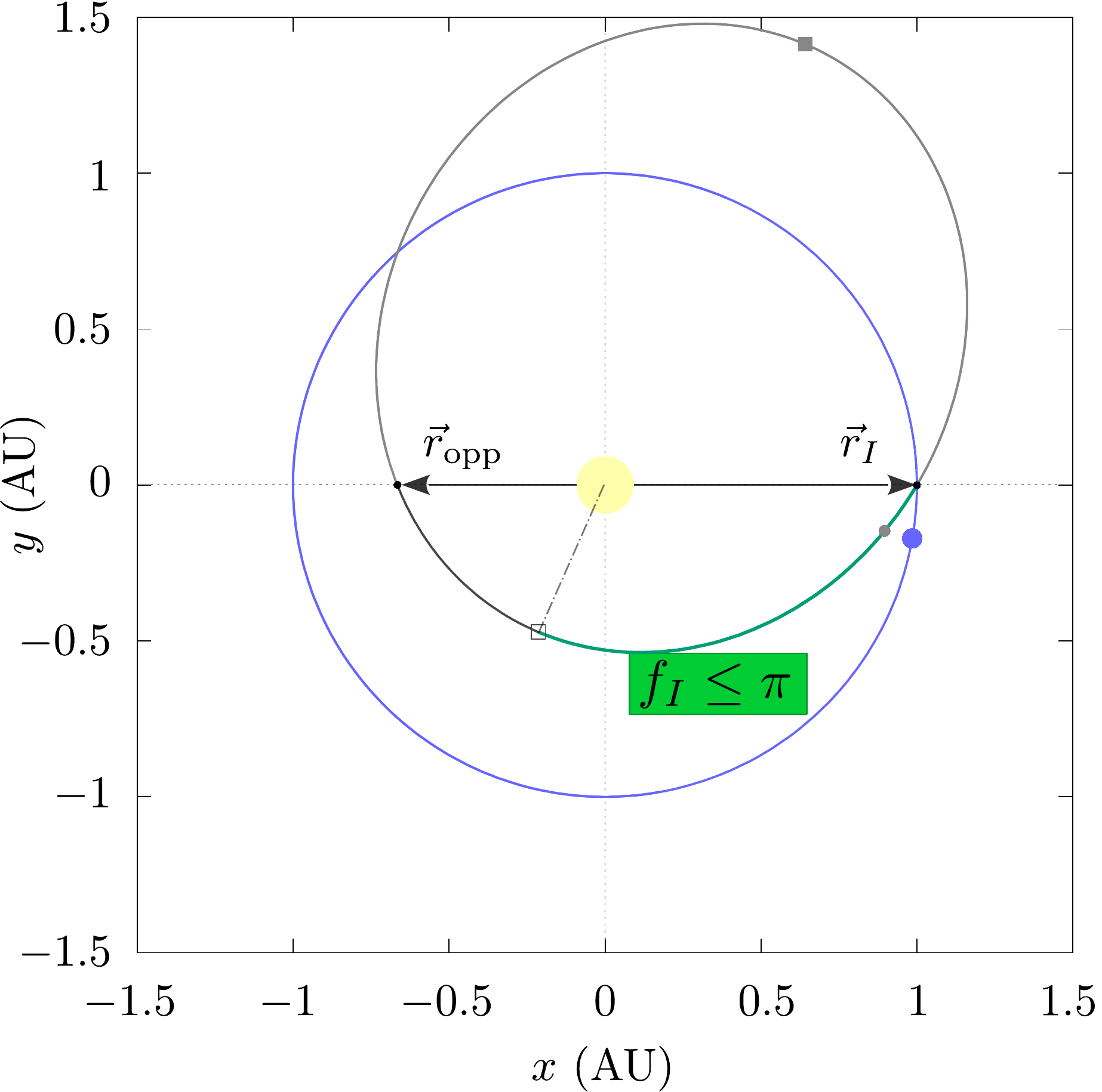} \hspace{.2cm} \includegraphics[height=7.cm]{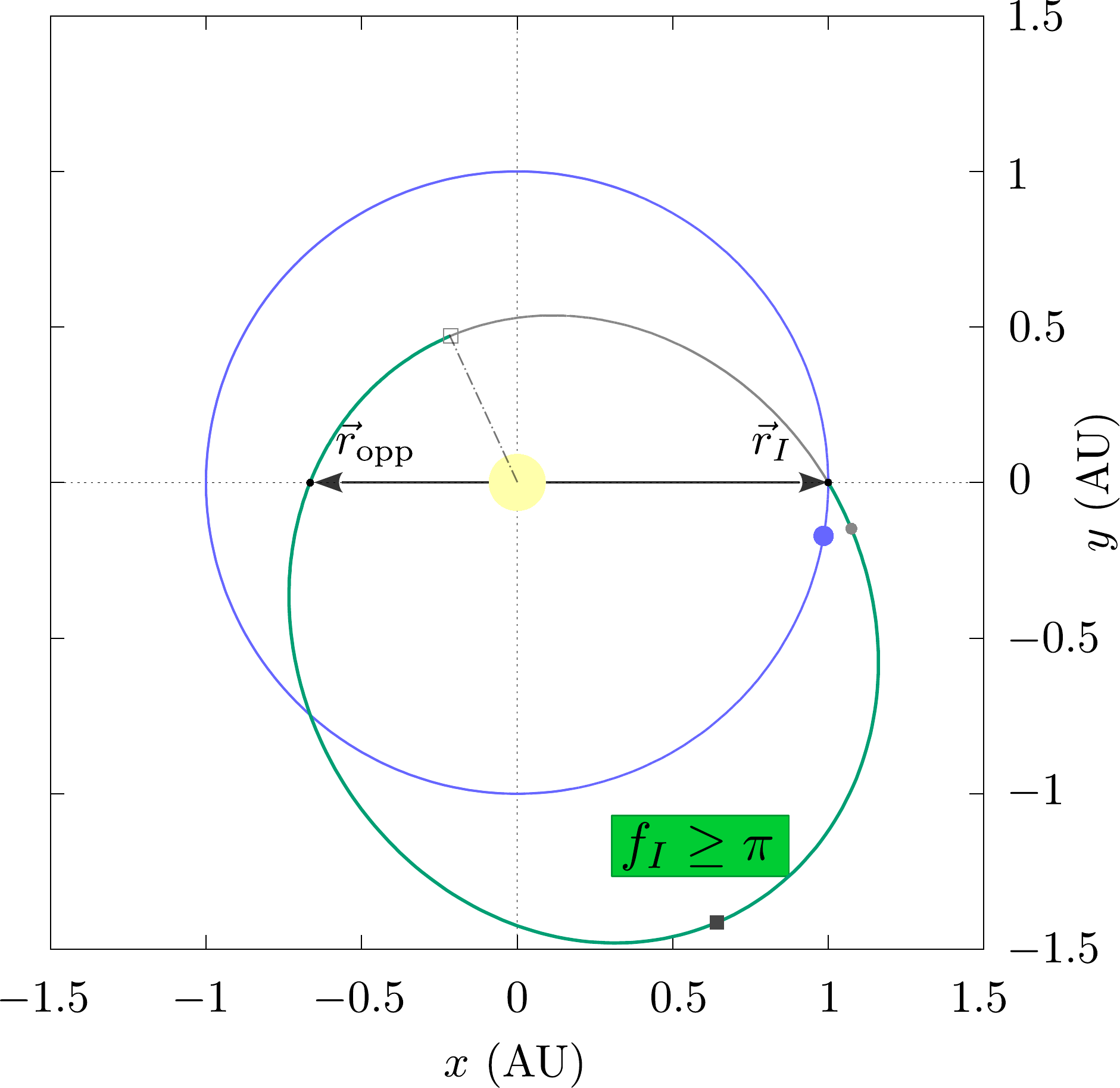}
        \caption{\small{Ecliptic projections: daytime impact (\emph{left}) and nighttime impact (\emph{right}). The circular Earth orbit is shown in blue; the 2 PHA orbits in grey are described by the same set $(P,e,i)$\hspace{1pt}---\hspace{1pt}same example as in Ref.~\cite{FB:2015}. Empty (resp.\@ full) squares indicate the perihelion (resp.\@ aphelion); the nodes of the PHA orbits, and the bodies location 10 days before impact are also shown (note: these PHAs are prograde, $i < 90$~deg).
	}}
        \label{fig:daytimenighttime}
    \end{center}
\end{figure}

From the perspective of any Earth-crossing asteroid, a strict impact with our planet must necessarily happen at a point on its heliocentric orbit such that, as trivially follows from the conic equation:
\begin{equation}
	r(f = f_I) = \frac{a(1 - {e}^2)}{1 + e \cos f_I} = 1~{\rm AU}, 	\label{eq:impactcondition}
\end{equation}
where $f_I\in [0,2\pi]$ is the asteroid true anomaly at impact.

Interestingly, we find that all the issues discussed in the previous section elegantly disappear if we use the parameter $f_I$ defined in Eq.~\eqref{eq:impactcondition}, for distinguishing daytime ($f_I \leq \pi$) and nighttime impacts ($f_I \geq \pi$); see again Fig.~\ref{fig:daytimenighttime}. Notice in passing that the two corresponding $f_I$, for the same set of semi-major axis and eccentricity, are such that their sum is obviously always $2\pi$, and that $|\pi - f_I|$ is the same for both.
This classification is not only simple, it remains valid in all cases. 	The asteroid true anomaly at impact is moreover not only truly geometrically intrinsic to the PHA orbit\hspace{1pt}---\hspace{1pt}depending only on $a$ and $e$\hspace{1pt}---\hspace{1pt}but also to the impact problem, since it simply tells where the impact takes place on the asteroid orbit. For comparison, the very concept of $\omega$ only makes sense when orienting the PHA orbit in three dimensions with respect to that of the Earth using as reference a rather arbitrary location (since the impact may happen there or not), which is moreover ill-defined when $i=0$.

Note that, even though the problem is independent of such a choice, it remains very simple to specify which node is impacted, for a given $f_I$.
The argument of perihelion of the PHA orbit $\omega$ is recovered once we specify whether the impact happens at the ascending or at the descending node of the PHA orbit. If $\vec{r}_I$ corresponds to the ascending node location, $\omega = 2\pi - f_I$ (and $\Omega = 0$ if we enforce Eq.~\eqref{eq:rI}); while $\omega = \pi - f_I$ (and $\Omega = \pi$), if it is instead the descending node.

\bigskip

Now, since it constitutes an appealing location for a possible deflection with a ballistic kinetic impactor, we are also interested in the other node of the PHA orbit. 
Given the necessarily limited launcher performance, possible mitigation missions with a ballistic trajectory are indeed bound to remain relatively close to the ecliptic plane: we cannot venture far out, especially not at large inclinations. For such cases, the optimal deflection point can therefore be expected to be found in the vicinity of the nodes of the PHA orbit, as was relied upon to determine initial guesses in Ref.~\cite{FB:2015}. For more modest inclinations, these initial guesses can at least provide a first reasonable option.

Here again, we do not want to have to make an explicit choice of whether this opposite node $\vec{r}_{\rm opp}$ corresponds to the ascending or to the descending node of the PHA orbit.
To get a general expression for the exact position of the opposite node, simply start from the following observation in terms of the true anomalies at the impact ($f_I$) and at the opposite node ($f_{\rm opp}$):
\begin{equation}
	f_{\rm asc.} \equiv 2\pi - \omega, \quad\textrm{while}\quad f_{\rm des.} \equiv \pi - \omega = f_{\rm asc.} - \pi,\notag
	\qquad
	\Rightarrow
	\qquad
	(f_I - f_{\rm opp}) \mod 2\pi = \pi = (f_{\rm opp} - f_I) \mod 2\pi
\end{equation}	
so that, using the conic equation, it is clear that
\begin{equation}
	\frac{|\vec{r}_{\rm opp}|}{|\vec{r}_I|} = \frac{1 + e \cos f_I}{1 - e \cos f_I},
	\qquad \textrm{while }
	\vec{r}_{\rm opp} = - |\vec{r}_{\rm opp}| \frac{\vec{r}_I}{|\vec{r}_I|},
	\label{eq:ropp}
\end{equation}
independently on whether the impact point corresponds to the ascending or descending node of the PHA orbit, allowing the discussion to remain completely general in the following.
At the same time, this of course also clearly identifies which of the two nodes $\vec{r}_I$ and $\vec{r}_{\rm opp}$ is closest to perihelion.

	\begin{figure}[h!!]
	\begin{center}
		\includegraphics[width = .5\textwidth]{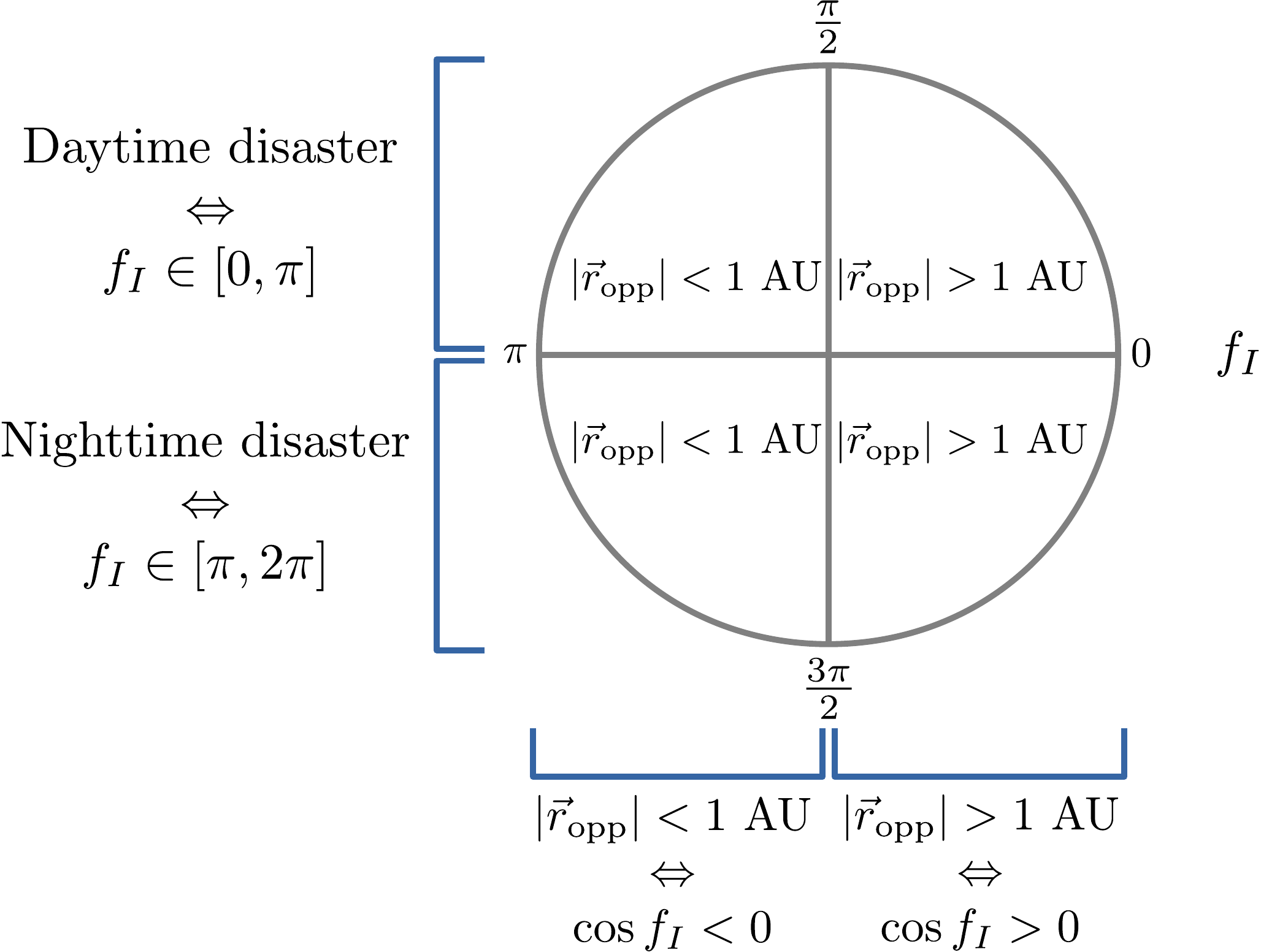}

	\end{center}
		\caption{Unit-circle representation of the asteroid true anomaly at impact $f_I$, showing how different properties are changing from one quadrant to the next.}
		\label{fig:scenarios_fI}
	\end{figure}

As Fig.~\ref{fig:scenarios_fI} schematically summarises, the various combinations of either daytime or nighttime impacts, with an opposite node either inside or outside of the Earth orbit, are then conveniently found to be separated in quadrants, giving a classification which always holds.
Note that, while still organised in quadrants, an equivalent figure for $\omega$ would instead strongly depend on whether the impact point is the ascending node or not, the properties of each quadrant then changing greatly depending on the case.

\subsection{Full description of any PHA orbit via its true anomaly at impact, eccentricity, and inclination}

While the asteroid true anomaly at impact $f_I$ has already proved to be useful, it is even more so once we go a step further, using  it to replace the semi-major axis. 
One can simply invert Eq.~\eqref{eq:impactcondition}, to obtain \mbox{$a = a(f_I,e)$}.
Generating the full parameter space of strictly impacting PHA orbits\footnote{Technically, since our only requirement is to consider Earth-crossing elliptic orbits around the Sun, comets are also included.} then becomes absolutely trivial, as it conveniently becomes a region without gaps, given by bounded dimensionless parameters:
\begin{equation}
	\textrm{any } (f_I, e, i),
	\label{eq:phaparam_fIei}
\end{equation}
where the asteroid true anomaly at impact $f_I$ is between $0$ and $2\pi$, the eccentricity of the asteroid orbit $e$ is between $0$ and $1$, and its inclination $i$ is between $0$ and $180$ deg.
There is a one-to-one correspondence: all such orbits correspond to fictitious PHAs, and all conceivable fictitious PHAs are contained in Eq.~\eqref{eq:phaparam_fIei}\hspace{1pt}---\hspace{1pt}and if needed, all the orbital elements of any such asteroid orbit can be straightforwardly recovered.

In comparison, with a more conventional parametrisation by means of the orbital period $P$, for a strict impact with a circular Earth orbit to be at all possible, a fictitious PHA must not only have an orbit such that (necessary but not sufficient condition):
 	\begin{equation}
		e \geq \left| 1 - { \left( \frac{P}{1~{\rm yr}} \right) }^{-\frac{2}{3}} \right|;
		\label{eq:borderPHAs}
	\end{equation}
	one should anyway still enforce the impact condition on $\omega$ as in Ref.~\cite{FB:2015}, or on $f_I$ as done here via Eq.~\eqref{eq:impactcondition}, and would have to deal separately with the two distinct branches (daytime and nighttime).

As a side note: since they physically describe the same thing, the two parametrisations are obviously connected. When given the semi-major axis and eccentricity of an orbit that strictly crosses the Earth orbit, one can compute the corresponding asteroid true anomaly at impact:
\begin{equation}
	f_I = \mathrm{acos}\left( \frac{1}{e} \left(\frac{a}{{\textrm {\scriptsize 1\hspace{2pt}AU}}} (1 - {e}^2) - 1\right) \right)
	\quad {\rm or}	\quad
	f_I = -\mathrm{acos}\left( \frac{1}{e} \left(\frac{a}{{\textrm {\scriptsize 1\hspace{2pt}AU}}} (1 - {e}^2) - 1\right) \right) + 2\pi.
	\label{eq:fI_of_a}
\end{equation}
As discussed in Sec.~\ref{sec:branches}, whenever using $a$ and $e$, the ambiguity between daytime and nighttime impact necessarily remain, which is why there must be two possible values for $f_I$ in Eq.~\eqref{eq:fI_of_a} when only that is provided. It can be lifted for instance by checking which of the two solutions leads to an impact using Eq.~\eqref{eq:impactcondition}, or with additional information: which branch it corresponds to; or with $\omega$, when knowing at which node the impact happens.

\subsection{Enabling an analytical study}

The new parametrisation presents several advantages, one of which being that it enables studying the problem analytically.
Relations and properties indeed become mathematically much simpler once expressed in terms of $(f_I,e,i)$, while preserving a clear meaning since the parameters themselves still represent physical concepts.

For instance, setting ourselves in the ecliptic frame as defined in Sec.~\ref{sec:intro}, in all generality the position of any impacting PHA, anywhere on its orbit, can in fact be written in a very compact way:
	\begin{equation}
		\vec{r}(f)  =  r \cos(f - f_I) \vec{e}_x + r \sin(f - f_I)\big[\cos i\ \vec{e}_y	\pm \sin i\ \vec{e}_z\big],
	\end{equation}
where $r=r(a(f_I,e),e,f)$ is given by the conic equation, and where the positive (resp.\@ negative) sign along $z$ corresponds to an impact at the ascending (resp.\@ descending) node of the PHA orbit\hspace{1pt}---\hspace{1pt}unimportant for the relative geometry and motions, as discussed earlier.
From there, one can express the polar unit vectors
	$\vec{e}_r(f) = \frac{\partial}{\partial r}\vec{r}$ and $\vec{e}_f(f) = \frac{1}{r}\frac{\partial}{\partial f}\vec{r}$: 	\begin{equation}
		\begin{split}
		\vec{e}_r(f) &=  \cos(f - f_I) \vec{e}_x + \sin(f - f_I)\big[\cos i\ \vec{e}_y	\pm \sin i\ \vec{e}_z\big]\\
	\textrm{and }	\vec{e}_f(f) &=  \cos(f - f_I)\big[\cos i\ \vec{e}_y	\pm \sin i\ \vec{e}_z\big] -\sin(f - f_I) \vec{e}_x.
		\end{split}
		\label{eq:e_r(f)__and__e_f(f)}
	\end{equation}
Finally, using these, the PHA heliocentric velocity at the impact location, expressed in the ecliptic frame $\vec{v}_{A,I} = v_{A,I,x} \ \vec{e}_x + v_{A,I,y} \ \vec{e}_y + v_{A,I,z} \ \vec{e}_z$, and the one at the opposite node $\vec{v}_{A,{\rm opp}} = v_{A,{\rm opp},x} \ \vec{e}_x + v_{A,{\rm opp},y} \ \vec{e}_y + v_{A,{\rm opp},z} \ \vec{e}_z$ in turn read
\begin{equation}
		v_{A,I,x} =  \sqrt{\frac{\mu_{\rm Sun}}{\rm 1\hspace{2pt}AU}} \frac{e \sin(f_I)}{\sqrt{ 1 + e \cos(f_I)}} = v_{A,{\rm opp},x} 
\end{equation}
\begin{equation}
v_{A,I,y} =  \sqrt{\frac{\mu_{\rm Sun}}{\rm 1\hspace{2pt}AU}} \sqrt{1 + e \cos(f_I)} \cos(i) \qquad
\textrm{and}
\qquad v_{A,{\rm opp},y} = -\sqrt{\frac{\mu_{\rm Sun}}{\rm 1\hspace{2pt}AU}} \frac{1 - e \cos(f_I)}{ \sqrt{ 1 + e \cos(f_I) } } \cos(i)
\end{equation}
\begin{equation}
v_{A,I,z} =  \pm\sqrt{\frac{\mu_{\rm Sun}}{\rm 1\hspace{2pt}AU}} \sqrt{1 + e \cos(f_I)} \sin(i)
\qquad
\textrm{and}
\qquad
v_{A,{\rm opp},z} = \mp\sqrt{\frac{\mu_{\rm Sun}}{\rm 1\hspace{2pt}AU}} \frac{1 - e \cos(f_I)}{ \sqrt{ 1 + e \cos(f_I) } } \sin(i),~\!
\label{eq:vel_z}
\end{equation}
where $\mu_{\rm Sun}$ stands for the gravitational parameter of the Sun.
For studying missions with a deflection taking place at either node of any general PHA orbit analytically, such relations turn out to be particularly useful.
As we are bound to remain concise here, only a few selected results can unfortunately be shown; the reader is referred to Ref.~\cite{AP:param} for more discussions and details. 

\subsection{Polar-plot representation of the PHA parameter space}

\begin{figure}[h!!]
    \begin{center}
        \includegraphics[height=6.25cm]{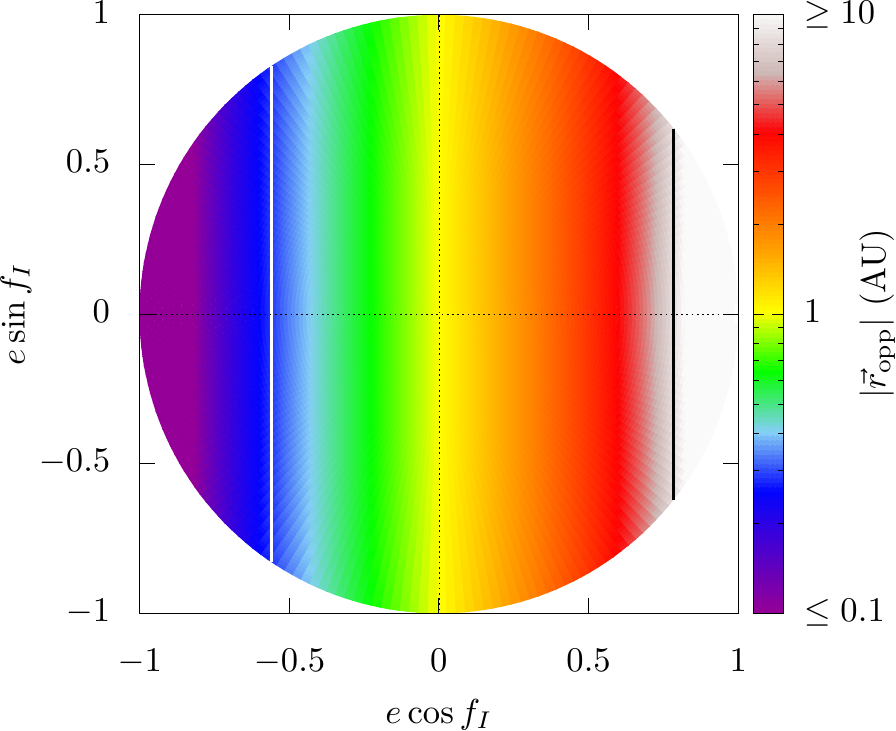}
        \caption{\small{Opposite-node location for the full PHA parameter space (independent of the inclination). The white/black curves coincide with the smallest/largest $|\vec{r}_{\rm opp}|$ that could be reached with $|\vec{v}_{\infty {\rm SC},L}| = 10$~km/s.}}
        \label{fig:ropp_POLAR}
    \end{center}
\end{figure}

Now, when using the new $(f_I,e,i)$ parametrisation, one thing that quickly becomes apparent is that most of the relevant quantities related to the deflection or the impact can actually be written as simple functions of $e\cos f_I$ and $e \sin f_I$. This suggests that a polar representation of the parametrisation could provide a complementary insight. In such plots, the eccentricity (going from $0$ to $1$) plays the role of the radial variable, while $f_I$ (going from $0$ to $2\pi$) is the angular variable; the inclination is fixed. The centre corresponds to a circular orbit (necessarily of radius $1$~AU), and the limit $e\rightarrow1$ would correspond to parabolas with different orientations ($f_I$).
A further advantage of this representation is that any possible PHA orbit uniquely corresponds to a single point: each point is a unique orbit, and all orbits are shown.

Figure~\ref{fig:ropp_POLAR} illustrates Eq.~\eqref{eq:ropp}. The advantages of the polar plot are self-evident here since $|\vec{r}_{\rm opp}|$ is only function of $e \cos f_I$; the dependency is clearly highlighted (it is also interesting to compare it with Fig.~\ref{fig:scenarios_fI}).
Together with $|\vec{r}_{\rm opp}|$, we moreover show two curves which appear as vertical lines on such a figure. These correspond to the matching with $|\vec{r}_{\rm opp}|$ of the smallest perihelion and largest aphelion that could be reached ballistically, given some maximal hyperbolic excess velocity at launch that we might want to consider.
Here, a launcher similar to the SLS Block 1B 8.4-m Fairing + EUS~\cite{SLS:2014} was assumed for the illustration (choosing $\max(|\vec{v}_{\infty{\rm SC},L}|) = 10$~km/s), but analytical expressions can be used for any different choice.
Such a comparison is particularly important in order to know what to expect from the feasibility point of view: knowing whether a deflection at the opposite node is an option or not, assuming a given launcher performance.
It is not possible to make a ballistic transfer to $\vec{r}_{\rm opp}$ for PHAs to the left (resp.\@ right) of the white (resp.\@ black) line.
This is actually going to be important for a crucial feasibility aspect of successful mitigation missions with kinetic impactors: the illumination conditions at deflection, notably the solar aspect angle $\phi_{\rm Sun}$:
	\begin{equation}
		\phi_{\rm Sun} \equiv \mathrm{acos}\left(\frac{-\vec{r}_{{\rm SC},D} \cdot \vec{v}_{\infty {\rm SC},D}}{ |\vec{r}_{{\rm SC},D}| \ |\vec{v}_{\infty {\rm SC},D}| } \right), \qquad  \textrm{with } \vec{v}_{\infty {\rm SC},D} = \vec{v}_{{\rm SC},D} - \vec{v}_{A,D},
		\label{eq:solaraspectangle}
	\end{equation}
	with $\vec{r}_{{\rm SC},D}$ and $\vec{v}_{{\rm SC},D}$, respectively the heliocentric position and velocity of the spacecraft at deflection.
Indeed, when the spacecraft is approaching the asteroid before impact, it should not have to look in a direction too close to the Sun, since it could saturate its on-board tracking systems and compromise the navigation, and therefore the deflection itself\hspace{1pt}---\hspace{1pt}as in Ref.~\cite{FB:2015} we shall then restrict it to be above a certain threshold.
From geometry, we can foresee that nighttime (resp.\@ daytime) impacts favour deflections in the vicinity of $\vec{r}_I$ (resp.\@ $\vec{r}_{\rm opp}$) when only considering $\phi_{\rm Sun}$, since the illumination conditions at deflection are clearly suitable there. The feasibility of transfers to $\vec{r}_{\rm opp}$ is then especially important for daytime PHAs.

\begin{figure}[h!!!]
    \begin{center}
        \includegraphics[height=6.25cm]{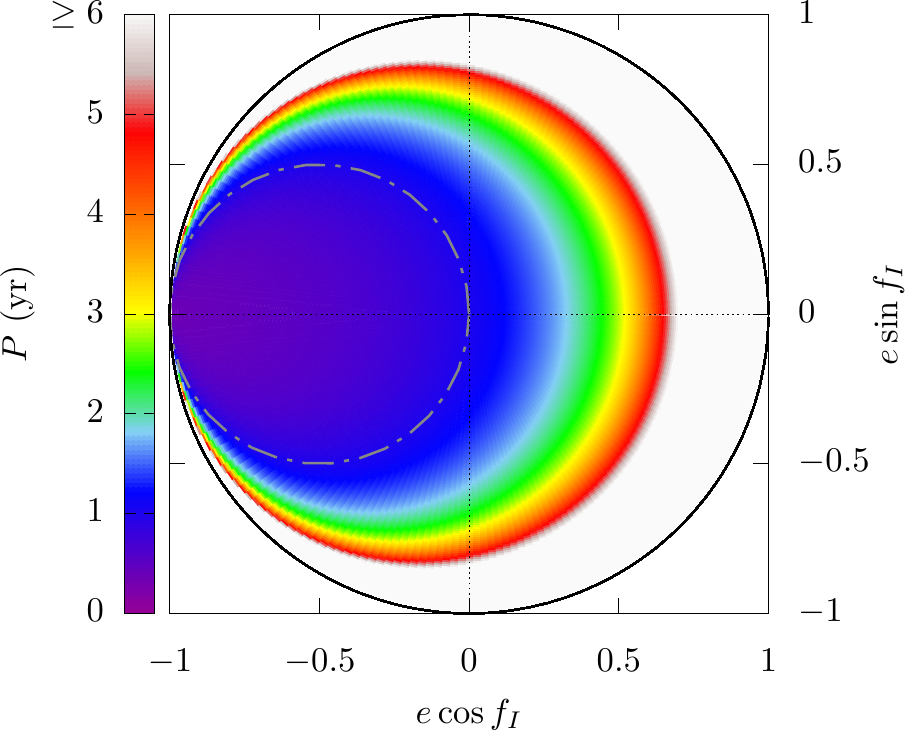} \hspace{.2cm} \includegraphics[height=6.25cm]{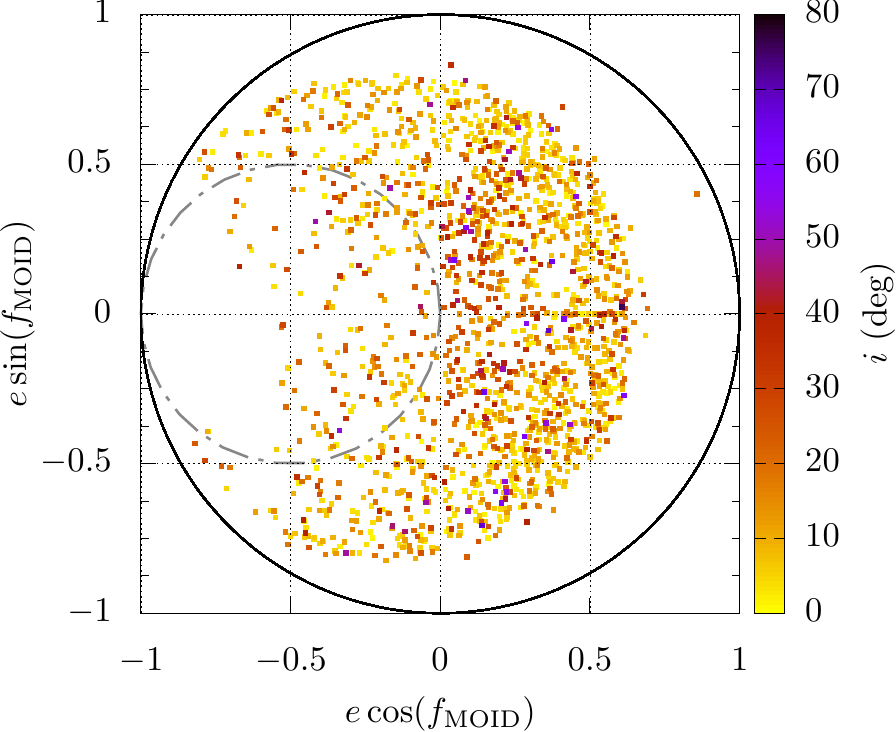}
        \caption{\small{Polar plots: orbital period (\emph{left}); actual asteroids from the PHA database of the Minor Planet Center, where the asteroid true anomaly at MOID is taken as a proxy for $f_I$ (\emph{right}). For strict impacts, Atens ($P < 1$~yr) are found inside the grey dashed circle corresponding to $P = 1$~yr, and Apollos ($P > 1$~yr), outside.}}
        \label{fig:Period_and_DB__POLAR}
    \end{center}
\end{figure}

Finally, to develop a better feel for polar plots, we can relate to Fig.~\ref{fig:Period_and_DB__POLAR}.  
On the left we plot the orbital period of each PHA orbit. The colour code stops at $P \geq 6$~yr, but that is for readability only: the full PHA parameter space really covers the whole disk. The boundary between Atens and Apollos is shown, and the symmetry between daytime and nighttime orbits, mentioned in Sec.~\ref{sec:introducing_fI} is clear.
To produce the plot on the right of that figure, we used the \texttt{pha\_extended} JSON database provided by the International Astronomical Union's Minor Planet Center (IAU MPC)\footnote{\texttt{\url{http://www.minorplanetcenter.net/data}}, accessed in early October 2016.}. Since no known PHA is foreseen to impact the Earth in the near future, we use the asteroid true anomaly at MOID as a proxy for $f_I$: both indeed coincide in the limit of MOID going to zero. One can see for instance that most PHAs are Apollos, that the eccentricity can be quite large, and the inclination, sizeable (its median is close to 10~deg). The period tends to be less than about $6$~yr, but it can sometimes be much larger.

\section{Deflection with ballistic kinetic impactors}

\subsection{Corresponding $B$-plane displacement}

For the deflection part, we do not strongly depart from the approach taken in Ref.~\cite{FB:2015}  and shall similarly only consider kinetic impactors on ballistic trajectories: no fly-by, no manoeuvres. While being restrictive, doing so provides a rather robust preliminary mission design, which could then be improved upon; moreover, limiting the scope is what facilitates or even enables studying all the PHAs at once.

Kinetic impactor themselves are typically not going to strongly alter the PHA orbital elements in the cases that we consider; the main outcome of the momentum transfer $\Delta\vec{p}_{{\rm SC},D}$ is then assumed to be a small effect on the orbital period of the PHA orbit: $\Delta P$.
The deflection is always considered to take place sufficiently in advance, letting the small effect on the orbital period enough time to accumulate.

A difference with Ref.~\cite{FB:2015}, is that for a mitigation mission (with a deflection at an epoch $T_D$), the accumulated along-track displacement when the asteroid finally approaches the Earth several years later (at an epoch $T_I$) should in fact read:
\begin{equation}
	\Delta s \approx - |\vec{v}_{A, I}| \Delta P \frac{T_I - T_D}{P}.\label{eq:corrected_FB_Delta_s}
\end{equation}
Accelerating a PHA at deflection will increase its orbital period, so that it will actually be late at its impact rendez-vous; while decelerating it will reduce its orbital period, meaning that the PHA will pass the impact point earlier than if it had not been deflected.
Right after deflection, of course, an accelerated PHA will move faster along track, and a decelerated one will move slower, but the effect on the period will soon take over. This is also discussed {\it e.g.} in Ref.~\cite{DachwaldKahleWie:2006}.

Note that Eq.~\eqref{eq:corrected_FB_Delta_s}, which neglects small non-secular effects~\cite{Izzo:2007}, is actually not conservative: compared with the secular part, the small offset induced can indeed have a different relative sign. Since we only consider long drifting time ($T_I - T_D$)
and moreover wish to enable a study that would ignore the phasing, here we shall neglect this small correction. To include it, see {\it e.g.} Refs.~\cite{VasileColombo:2008,BombardelliBau:2012}.

\begin{figure}[h!]
    \begin{center}
        \includegraphics[width=0.35\textwidth]{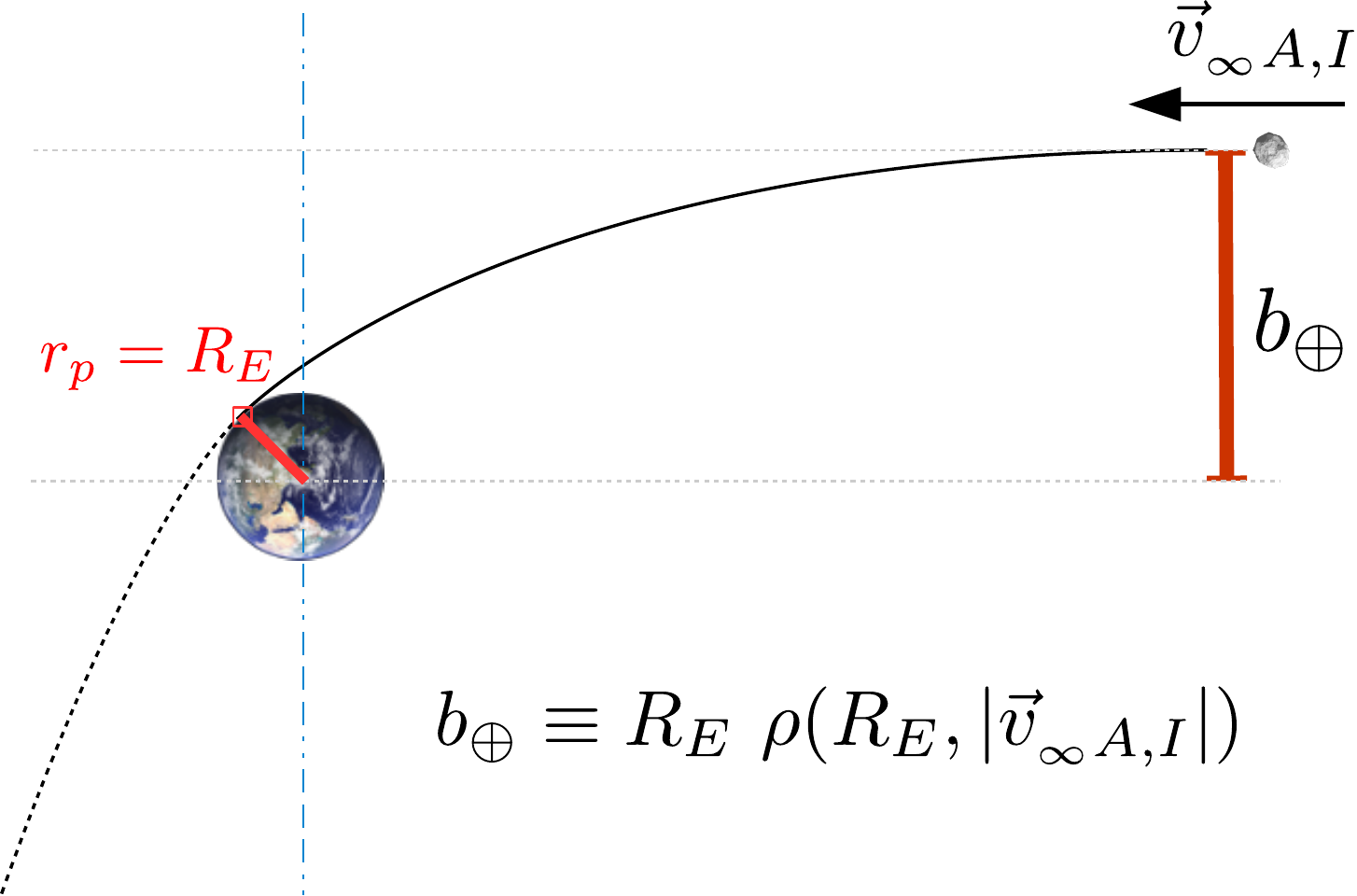}
        \caption{\small{For each PHA, $b_{\oplus}$ represents a characteristic length scale in the $B$-plane.}}
        \label{fig:boplus}
    \end{center}
\end{figure}

Since our aim is to remain quite general, the discussion is done in terms of $B$-plane displacement $\Delta B$, as in Ref.~\cite{FB:2015}. Doing so indeed allows one to present results that can be used for any impacting asteroid (not only those with vanishing impact parameters), without having to specify what exactly its $B$-plane impact parameter actually is. Instead, one can then discuss in terms of the worst possible case, which would require $\Delta B > 2 b_{\oplus}$~\cite{FB:2015}, where $b_{\oplus}$ is the impact parameter such that the perigee of the incoming hyperbola at close approach is given by the Earth radius $R_E$; see Fig.~\ref{fig:boplus}.
Assuming a two-body problem, the influence of the Earth gravity is indeed taken into account at close approach (focus effect), as often done in the literature ever since the seminal work of \"Opik, see {\it e.g.}~\cite{Oepik:1976}, via:
\begin{equation}
	\rho(r_p, |\vec{v}_{\infty A,I}|) = \sqrt{  1 + \frac{ 2 \mu_{E} }{ r_p \ {\big|\vec{v}_{\infty A,I}\big|}^2 }  }, \qquad \textrm{with } \vec{v}_{\infty A,I} = \vec{v}_{A,I} - \vec{v}_{E,I},
\end{equation}
with $\mu_{E}$, the Earth gravitational parameter; $\vec{v}_{E,I}$, the Earth heliocentric velocity at $\vec{r}_I$.

Due to the focus effect, the exact displacement required to avoid deflection is a function of $|\vec{v}_{\infty A,I}|$. Rather than adapting it from one PHA to another, using a fixed scale (such as kilometres or number of Earth radii), we use $b_{\oplus}$ as a scale with a physical meaning close to what we want to achieve and which inherently adapts itself.
While the length $b_{\oplus}$ will not be the same for every PHA, the deflection results given in units of $b_{\oplus}$ can be immediately compared among PHAs. What matters is indeed not so much the exact displacement in kilometres (which can of course be easily calculated); it is rather to measure whether the impact can be avoided\hspace{1pt}---\hspace{1pt}and using $b_{\oplus}$ as a unit of length provides a clear answer to this.

Finally, the linearised $B$-plane displacement $\Delta B =  \Delta s \sin \gamma$, where the projection on the $B$-plane is done via
\begin{equation}
	\sin \gamma = \sin\left( \mathrm{acos}\left( \frac{ \vec{v}_{A,I} \cdot \vec{v}_{\infty A,I} }{ |\vec{v}_{A, I}| \ |\vec{v}_{\infty A,I}| } \right) \right),
\end{equation}
can be estimated by (we conservatively assume $\beta = 1$, {\it i.e.\@} perfectly inelastic collisions)
\begin{equation}
	\Delta B 	\approx \underbrace{- |\vec{v}_{A, I}| \sin\gamma \frac{3a}{\mu_{\rm Sun}} \ \frac{\beta}{m_A}}_{\textrm{not/barely affected by the mission}} \ \underbrace{(\vec{v}_{A, D} \cdot \Delta\vec{p}_{{\rm SC},D} ) ( T_I - T_D )}_{\textrm{merit function $\mathbb{M}$}}. \label{eq:corrected_DeltaB}
\end{equation}

\subsection{Ignoring the phasing}

Maximising the $B$-plane displacement $|\Delta B|$ with a kinetic impactor requires designing a deflection mission which optimises the merit function $\mathbb{M}$~\cite{FB:2015}:
\begin{equation}
	\mathbb{M} = \underbrace{(\vec{v}_{A, D} \cdot \Delta\vec{p}_{{\rm SC},D})}_{\equiv\mathcal{M}} \ \underbrace{( T_I - T_D )}_{\equiv \Delta T},
\end{equation}
the other relevant quantities entering Eq.~\eqref{eq:corrected_DeltaB} being fixed by the PHA orbit and the associated relative geometry, as well as by the physical properties of the bodies. In particular, a kinetic impactor will not significantly alter the close-approach geometry.

For a study that ignores the phasing\hspace{1pt}---\hspace{1pt}which for instance allows identifying and understanding better what the best possible solution would be and why, if it was not for timing issues\hspace{1pt}---\hspace{1pt}, one can for instance take $\Delta T = T_I - T_D = 1$~yr and is then left with the optimisation of $\mathcal{M}$, which is essentially the geometric part of the merit function.
In such a study, one allows for all the possible geometries at deflection.

When ignoring the phasing, it is furthermore useful to consider the $B$-plane displacement per interval of time following the deflection, which we write $\Delta\mathcal{B}$ and call the $B$-plane yearly drift. It is simply obtained by replacing the merit function $\mathbb{M}$ by $\mathcal{M}$ in Eq.~\eqref{eq:corrected_DeltaB}; it will be given in units of $b_{\oplus}$/yr in what follows.

\subsection{Some optimisation results}

\begin{figure}[h!!!]
    \begin{center}
        \includegraphics[width=7.75cm, trim = 0 24 0 0 , clip]{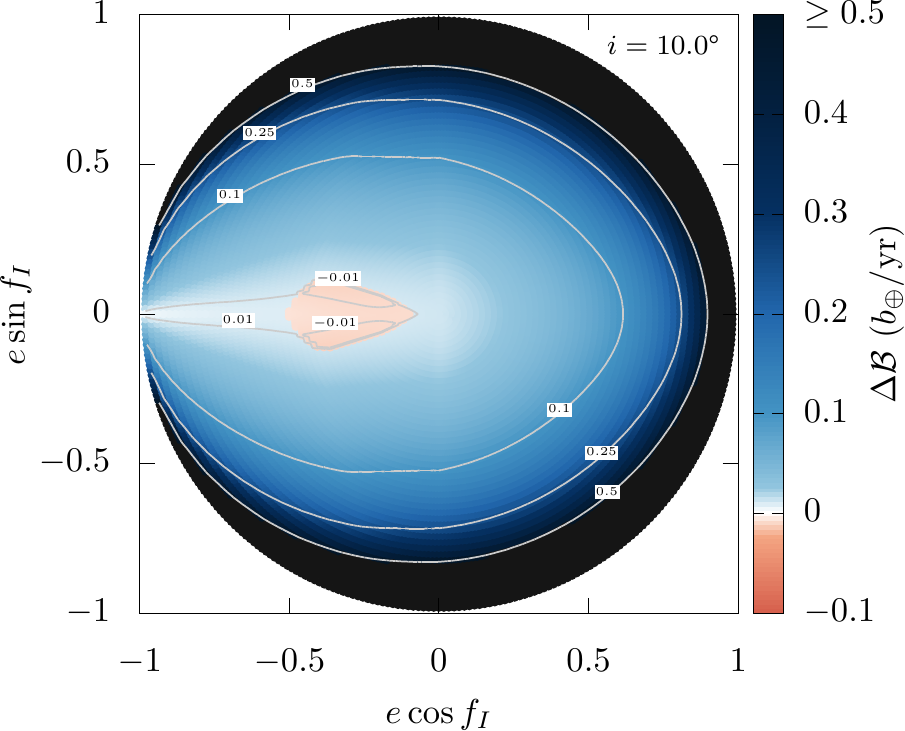} \hspace{.2cm} \includegraphics[width=7.75cm, trim = 0 24 0 0 , clip]{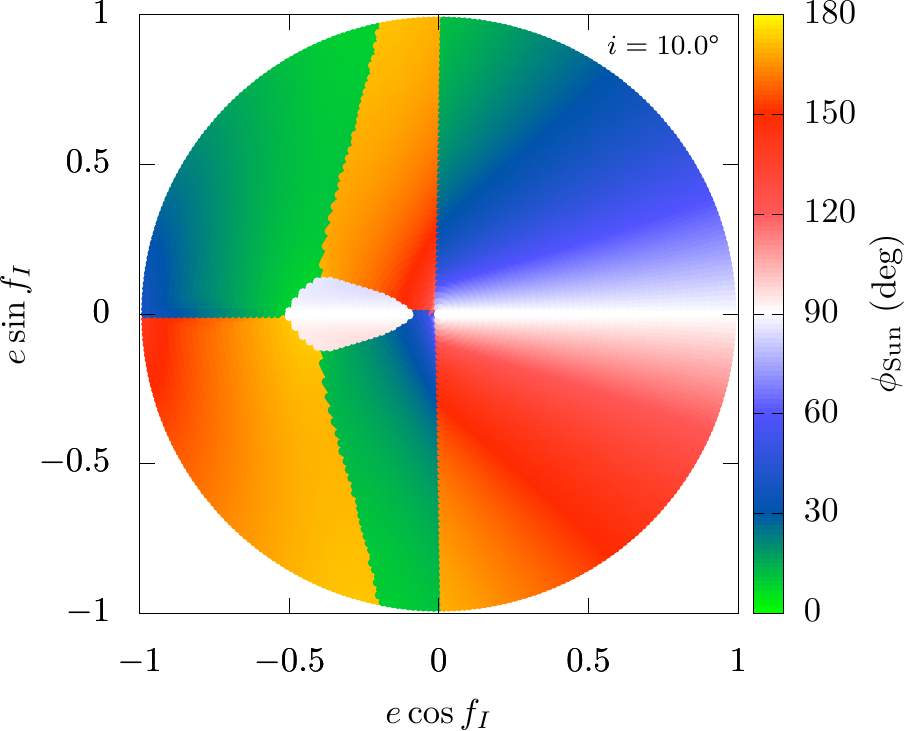}\\
        \includegraphics[width=7.75cm]{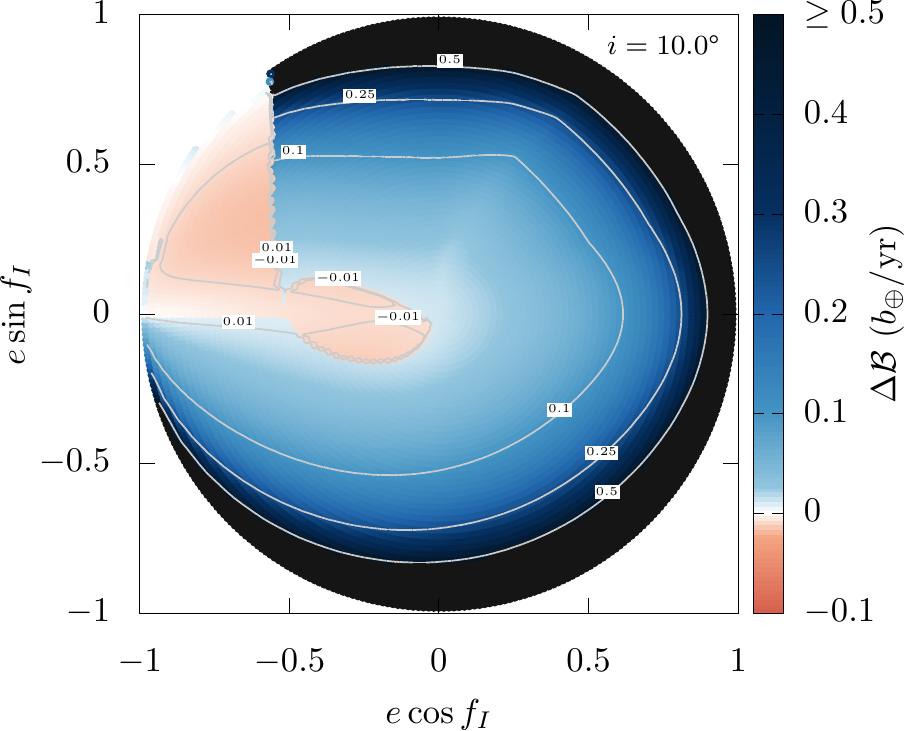} \hspace{.2cm} \includegraphics[width=7.75cm]{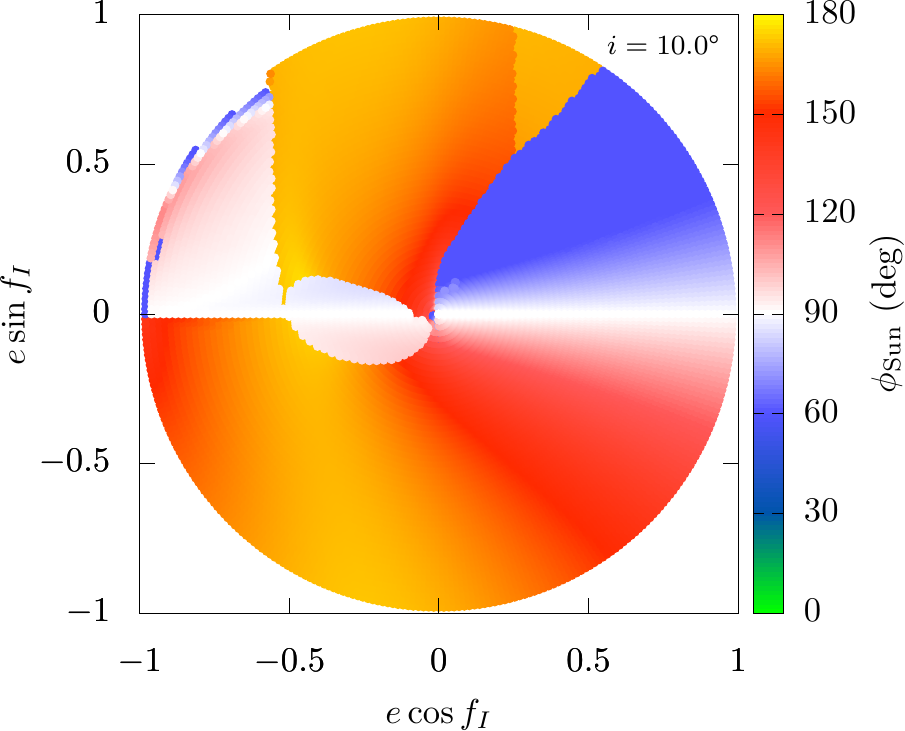}
        \caption{\small{Results of a numerical optimisation ignoring the phasing, in which the solar aspect angle at deflection is either totally unconstrained (\emph{top}) or required to be $\phi_{\rm Sun } \geq 60$~deg (\emph{bottom}).}}
        \label{fig:nor_best_of_best_incl10deg}
    \end{center}
\end{figure}

Figure~\ref{fig:nor_best_of_best_incl10deg} shows for instance the final results of a numerical optimisation ignoring the phasing for PHA orbits with an inclination of 10 degrees ($\sim$ median inclination, see again Fig.~\ref{fig:Period_and_DB__POLAR}). It shows both the yearly drift $\Delta \mathcal{B}$ and the corresponding solar aspect angle at deflection $\phi_{\rm Sun}$ in two different settings: with and without a cut on $\phi_{\rm Sun}$. The launcher performance as well as the mass assumed for all PHAs are chosen here to match what was considered in Ref.~\cite{FB:2015}, to facilitate direct comparisons. These are respectively the SLS Block 1B (with $\max(|\vec{v}_{\infty{\rm SC},L}|) = 10$~km/s), and $m_A = 7.77 \times 10^7$~t ({\it i.e.\@} a diameter of about 400~m and a density of $2$~t/m$^3$); though of course one can easily rescale for another $m_A$ in Eq.~\eqref{eq:corrected_DeltaB}.
Here also, the effect of the declination at launch on the launcher performance was ignored.

Focusing on the deflection (left panels, yearly drift $\Delta \mathcal{B}$), notice for instance the acceleration region ($\Delta \mathcal{B}<0$) that appears when the solar aspect angle is required to be above a certain threshold. What determines its location is in fact directly linked to the feasibility discussion that was made about Fig.~\ref{fig:ropp_POLAR}. There are many interesting discussions that can be made, linking the different regions in those figures to analytical results, and of course also study the influence of the inclination. For more on this, including the differences for performing deflection missions to PHAs with either daytime or nighttime impact, the interested reader is referred to Ref.~\cite{APJS:defl}, which is entirely devoted to this subject.

\begin{figure}[h!]
    \begin{center}
	\includegraphics[height=7.cm]{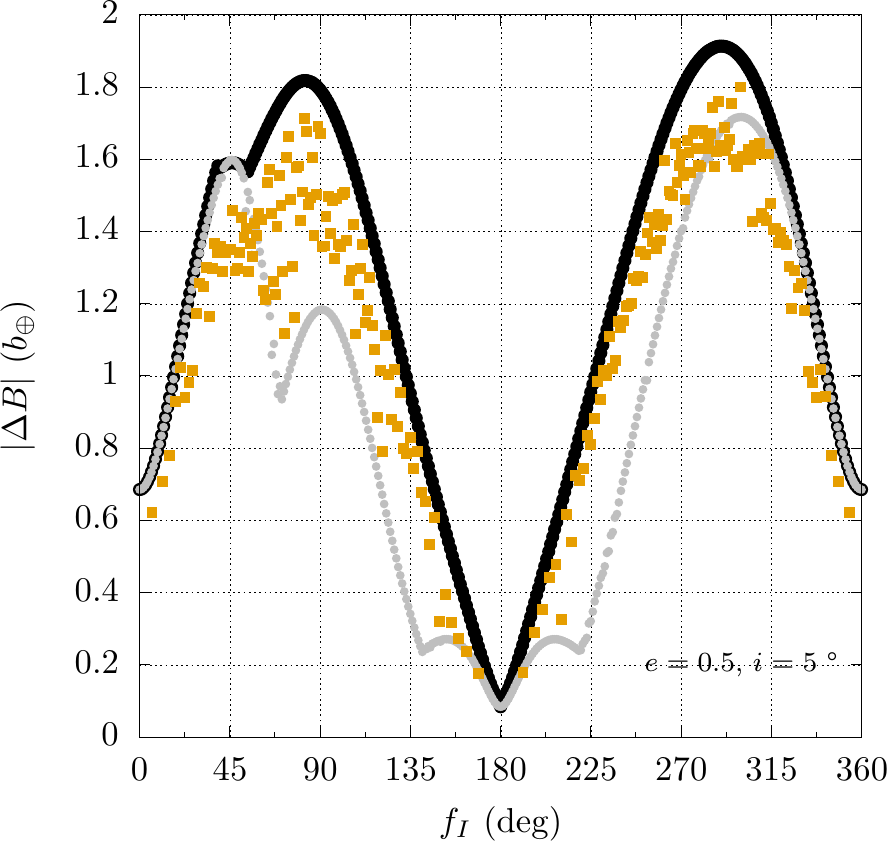}
	\includegraphics[height=7.cm, trim = 15 0 0 0 , clip]{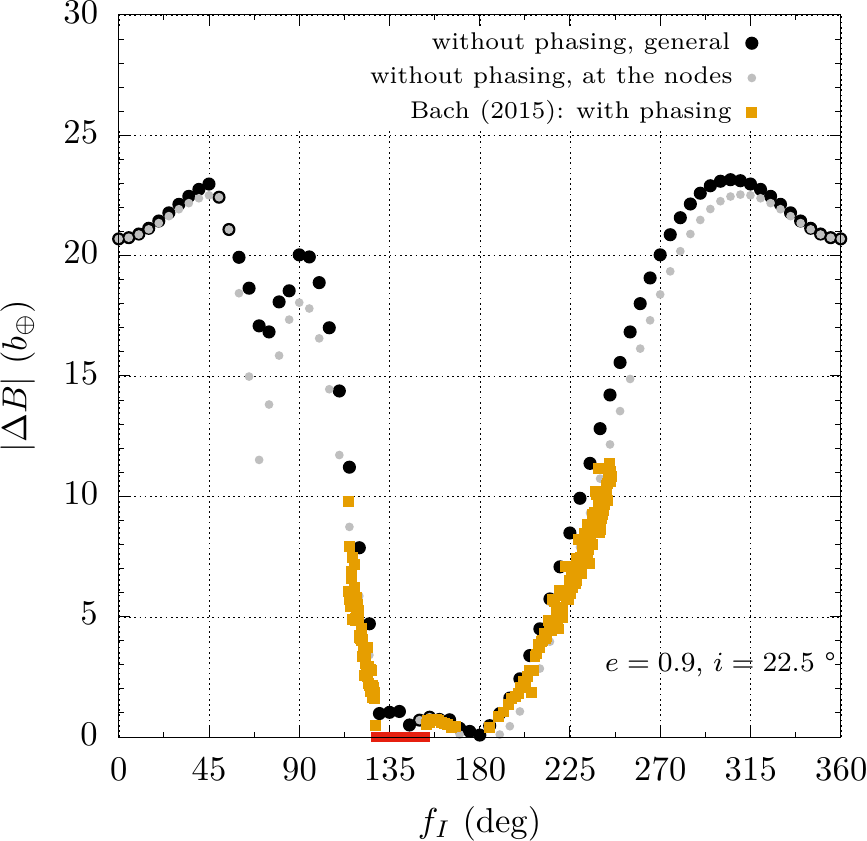}
        \caption{\small{Comparison with an optimisation with phasing~\cite{FB:2015} in 2 examples with a cut \mbox{$\phi_{\rm Sun} \geq 45$ deg.} PHAs have $P \leq 6$~yr in Ref.~\cite{FB:2015} and therefore do not cover the whole range of $f_I$ for $e = 0.9$ (\emph{right}).}}
        \label{fig:comparisonwithphasing}
    \end{center}
\end{figure}

Finally, we give a quick comparison with the results of an optimisation with phasing in Fig.~\ref{fig:comparisonwithphasing}. For that, taking the raw simulation data from Ref.~\cite{FB:2015}, we first make a transformation into our new parametrisation. We then show on the same plot these data and our own optimisation results without phasing for the same asteroids, where the $B$-plane displacement for the results without phasing is simply given by
\begin{equation}
	\Delta B = \Delta \mathcal{B} \times \Delta T,
\end{equation}
taking $\Delta T = 20$~yr here, since this was the warning time assumed in Ref.~\cite{FB:2015}.
As expected, these agree quite well, and even with the results for which the deflection was required to take place exactly at one of the nodes of the PHA orbit (given how crude that is). The results of the general optimisation without phasing of course lead to a larger change of the impact parameter, since it corresponds to the best possible deflection in a given framework. While not shown here, the analytical formulae for a deflection at the nodes (results of the analytical optimisation, assuming no cut on $\phi_{\rm Sun}$) would essentially be indistinguishable from those of the numerical optimisation restricted at the nodes with no solar-aspect-angle threshold. One benefit of these formulae is of course that one does not need to solve a single Lambert problem, and can straightaway choose the inclination and other parameters to its liking, simply providing a fit of the launcher performance.
Assuming a deflection happening strictly at the nodes is of course more satisfactory the larger the inclination (since deflecting there eventually becomes the only option); nonetheless, even at more modest inclinations, these results remain useful, at the very least since they allow for a quick assessment of what may be expected for asteroids in different regions of parameter space; including for the solar aspect angle at deflection, which is also derived analytically.

\section{Conclusion}

In this paper, a new parametrisation for Earth-crossing orbits was introduced and shown to be well-suited for studying strict impacts, since it preserves the symmetries of the impact problem.
The definition of the full PHA parameter space then becomes straightforward, corresponding in fact to a continuous and bounded region without gaps, clearly delimited via three dimensionless physical parameters.

A further benefit of this new parametrisation is to strongly highlight a number of properties and symmetries of the impact and deflection problems when presented in a polar plot\hspace{1pt}---\hspace{1pt}in which there is moreover a one-to-one correspondence between each point and each possible impacting orbit (no branches, nor redundancy).
It is also well-defined, even for PHA orbits with zero inclination.
Daytime and nighttime impacts are clearly separated and, for any orbit, the feasibility of a deflection in the vicinity of the opposite node with a given launcher is easily assessed. Equations also take a much simpler form, opening the possibility of studying the problem analytically.
Finally, all this comes with no loss of generality, except assuming a circular Earth orbit strictly crossed by elliptical trajectories.

Using the new parametrisation, mitigation missions by means of ballistic kinetic impactors were then studied, considering at once the complete PHA parameter space.
Analytical results for a deflection restricted at the nodes were derived, which, despite inherent limitations, give the flexibility of having formulae at hand both for the deflection and for the solar aspect angle; becoming more and more accurate at larger inclinations. An important part of this work is then a numerical optimisation study for different values of the PHA-orbit inclination, which ignores the phasing, allowing to identify the best possible yearly-drift that could be achieved with a ballistic kinetic impactor for any asteroid.
Taken together, these give both a qualitative understanding (mapping types of missions to asteroids) and quantitative results (best conceivable yearly drift).
Finally, those findings are found to be compatible and to complement what one would find when including the phasing, as they can give an insight of what makes the best possible mission, and allow among other things to summarise the general trends.

\section*{Acknowledgments}

It is with great pleasure that the authors thank Fabian Bach, who shared his raw numerical data, and allowed for their transformation and use for comparison purposes with the results obtained in this work.
The authors also gladly thank Michael Khan for a number of interesting discussions and comments.

This research has made use of data and/or services provided by the International Astronomical Union's Minor Planet Center.

\bibliographystyle{model1-num-names}   
\bibliography{references}             

\vspace{0.25in}

%
%

\end{document}